\documentclass[journal,onecolumn]{IEEEtran}
\usepackage{cite}
\usepackage{graphicx}
\usepackage{amssymb}
\usepackage[cmex10]{amsmath}

\usepackage{algorithm}
\usepackage{algpseudocode}%

\newcommand\floor[1]{\lfloor#1\rfloor}
\newcommand\ceil[1]{\lceil#1\rceil}
\newcommand\round[1]{\lceil#1\rfloor}
\DeclareMathOperator*{\argmax}{arg\,max}
\DeclareMathOperator*{\argmin}{arg\,min}
   
\newtheorem{theorem}{Theorem}

\newtheorem{defn}{Definition}
\newtheorem{rem}{Remark}

\begin{document}
%
\title{Polynomially Solvable Instances of the Shortest and Closest Vector Problems with Applications to Compute-and-Forward}

\author{Saeid~Sahraei,~\IEEEmembership{Student Member,~IEEE,}
         and~Michael~Gastpar,~\IEEEmembership{Fellow,~IEEE}

\thanks{This work has been supported in part by the European ERC
Starting Grant 259530-ComCom. }
\thanks{This work has been partially presented at the 52nd Allerton Conference on Communications, Control, and Computing.}
\thanks{S. Sahraei and M. Gastpar are with the school of Computer and Communication Sciences, {\'E}cole Polytechnique F{\'e}d{\'e}rale de Lausanne, Lausanne 1015, Switzerland (email: saeid.sahraei@epfl.ch; michael.gastpar@epfl.ch). }}
\maketitle

\begin{abstract}
A particular instance of the Shortest Vector Problem (SVP) appears in the context of Compute-and-Forward. Despite the NP-hardness of the SVP, we will show that this certain instance can be solved in complexity order $O(n\psi\log(n\psi))$ where $\psi = \sqrt{P\|{\bf h}\|^2+1}$ depends on the transmission power and the norm of the channel vector. We will then extend our results to Integer-Forcing and finally, introduce a more general class of lattices for which the SVP and the and the Closest Vector Problem (CVP) can be approximated within a constant factor.
\end{abstract}



\section{Introduction}
\IEEEPARstart{T}{he} shortest Vector Problem (SVP) is the problem of finding the shortest non-zero vector of a lattice. It can be mathematically expressed as:
\begin{equation}
{\bf a}^* = \argmin_{{\bf a}\in\mathbb{Z}^n\backslash\{\bf 0\}}\|{\bf Aa}\|^2 = \argmin_{{\bf a}\in\mathbb{Z}^n\backslash\{\bf 0\}}{\bf a}^T{\bf G}{\bf a}
\label{eqn:svp}
\end{equation}
where the full-rank matrix ${\bf A}\in \mathbb{R}^{n\times n}$ is the lattice basis and ${\bf G = A}^T{\bf A}$ is called the Gram matrix of the lattice. 
The Closest Vector Problem (CVP), is the problem of finding the closest vector of a lattice to ${\bf y}$, an arbitrary vector in $\mathbb{R}^n$: 
\begin{equation}
{\bf a}^* = \argmin_{{\bf a}\in\mathbb{Z}^n}\|{\bf Aa-y}\|^2 =  \argmin_{{\bf a}\in \mathbb{Z}^n} {\bf a}^T{\bf Ga} - 2{\bf y}^T{\bf Aa}+ {\bf y}^T{\bf y}.
\label{eqn:cvp}
\end{equation}
The CVP and the SVP are known to be NP-hard under randomized reduction\cite{micciancio2001hardness,micciancio2001shortest}. In fact, for a general lattice, there is not even an efficient constant-factor approximation algorithm known for these problems. Discovering such algorithms would have significant implications in terms of the hierarchy of complexity classes \cite{khot2004hardness,alekhnovich2005hardness,micciancio2001shortest}. Currently, the best known polynomial complexity approximation algorithms for the SVP/CVP only achieve exponential approximation factors \cite{lenstra1982factoring,gama2008finding}.

On the bright side, efficient algorithms for special lattices have been known for a long time. For instance Gauss found an algorithm for solving the SVP in dimension two. Conway in \cite{bannai1999sphere} provides exact algorithms for a class of root lattices in higher dimensions. Based on \cite{conway1992low} McKilliam \cite{mckilliam2012finding} showed that if an obtuse superbase for a lattice is known, the SVP and the CVP can be solved in polynomial complexity.

In this work, we introduce new classes of lattices where the SVP and the CVP are of polynomial complexity.
These classes are inspired by recently proposed cooperative communication strategies referred to as
``Compute-and-Forward" \cite{nazer2011compute} and ``Integer-Forcing" \cite{zhan2014integer}. Optimizing the computation rate achieved by these strategies involves solving particular instances of the SVP. We will first show that under certain conditions on the eigenvalues of the matrix ${\bf G}$ in Equations \eqref{eqn:svp} and \eqref{eqn:cvp}, the solution to the SVP and the CVP can be found in complexity order
\begin{equation}
O\left(n^{k+1}(2\ceil{\psi}+2)^{k+1}\right).
\end{equation}
We will then show that the instances of the SVP that we are interested in satisfy these conditions. As for Integer-Forcing, the parameters $k$ and $n$ respectively stand for the number of antennas at the receiver and the number of transmitters. Here $\psi = \sqrt{1 + P\gamma_{max}^2}$ depends on $P$, the transmission power and $\gamma_{max}^2$, the largest eigenvalue of ${\bf H}{\bf H}^T$ where ${\bf H}$ is the channel matrix.
For Compute-and-Forward the complexity can be further reduced to 
\begin{equation}
O(n\psi\log(n\psi))
\end{equation} 
where $\psi = \sqrt{P\|{\bf h}\|^2+1}$ and ${\bf h}$ is the channel vector.

We will then proceed by introducing a larger class of lattices for which a constant approximation factor for the SVP and the CVP can be found in polynomial complexity.

{ There is a sizable literature concerning the particular instance of the SVP that appears in Compute-and-Forward and Integer-Forcing. Modifications to Sphere Decoding and Schnorr-Euchner algorithms \cite{wen2014compute,wen2016efficient},  Quadratic Programming Relaxation \cite{zhou2014quadratic}, Branch and Bound algorithm \cite{richter2012efficient}, Slowest Descent method \cite{wei2013integer}, modifications to the LLL algorithm \cite{sakzad2013integer,lenstra1982factoring}, exhaustive search over a reduced search space \cite{hejazi2013simplified}, etc  are some of the approaches taken by the community. Some of these works have excellent (sometimes linear) {\it average}  complexity based on simulation results. Our work is distinguished in that we provide rigorous theoretical guarantees on both complexity (worst-case) and correctness of the algorithm. In addition, direct extensions of our work have been shown to demonstrate linearithmic average complexity \cite{wen2016linearithmic,huang2016low}. Our algorithm has further been extended to the Compute-and-Forward problem over rings of Gaussian integers and Eisenstein integers \cite{sun2013lattice,liu2016efficient}. Finally, another line of research concerns the recoverability of the messages at the receivers in a multi-relay scenario \cite{barreal2015low,richter2015non,zhou2016efficient},  where decoding the best equation by each relay node may not be the best approach.}

The paper will continue as follows. We will briefly discuss the Compute-and-Forward technique and its connection with the SVP in Section \ref{sec:IP1}. We will demonstrate how this particular instance of the SVP can be solved efficiently.  In Section \ref{sec:IPK} we will extend our results to Integer-Forcing. Next, in Section \ref{sec:almostIPK} we will show that the SVP and the CVP can be approximation up to a constant factor for a much larger class of lattices. Finally, in Section \ref{sec:reduction} we will discuss an open problem in the context of lattice reduction.


\section{Notation}
We use boldface capital letters for matrices and boldface lowercase letters for vectors. Scalars are represented by plain letters. All vectors are column vectors by default. All the vector inequalities used throughout the paper are elementwise. For a square matrix ${\bf A}$, the operator $diag({\bf A})$ returns a column vector which consists of the diagonal elements of ${\bf A}$. The operator $\ceil{\cdot}$ returns the smallest integer greater or equal to its input. The two operators $\lceil\cdot\rfloor$ and $\lfloor\cdot\rceil$ return the closest integer to their inputs. Their difference is at half-integers: the former rounds the half-integers up and the latter rounds them down. Once these operators are applied to a vector, they act element-wise. We use $\|\cdot\|$ to represent the 2-norm of a vector. For an $n\times n$ matrix ${\bf A}$ and for a set ${\pi \subseteq \{1,\dots,n\}}$ we define ${\bf A}_\pi$ as the submatrix of ${\bf A}$ which consists of the rows indexed in ${\pi}$. For a vector ${\bf a}\in \mathbb{R}^n$, we define ${\bf a}_\pi$ in a similar way. ${\bf I}$ is the identity matrix and ${\bf 1}$ and ${\bf 0}$ represent the all-one and all-zero vectors respectively. Finally, $\mathbb{R}$ is the set of real numbers and $\mathbb{Z}$ the set of integers.

\section{$\text{IP}^1$ Matrices and Compute-and-Forward}
\label{sec:IP1}

The initial motivation behind this work is the problem of maximizing the achievable computation rate of Compute-and-Forward. In this section, we will present a short introduction to this relaying technique and establish its connection with the SVP. We will then demonstrate how this particular instance of the SVP can be solved in polynomial complexity.\subsection{Compute-and-Forward}
Compute-and-Forward is an emerging relaying technique in wireless multiuser networks. Contrary to the conventional approaches, Compute-and-Forward does not view interference as inherently undesirable. The key idea is to recover integer linear combinations of transmitted codewords as opposed to decoding individual transmitted messages. Nested lattice codes ensure that these integer linear combinations are codewords themselves. Compute-and-Forward has the potential to increase the achievable rate compared to the traditional relaying techniques, as the analysis suggests in \cite{nazer2011compute,nam2008capacity,wilson2010joint}.

\begin{figure}[h]
\begin{center}
\includegraphics[width=8cm]{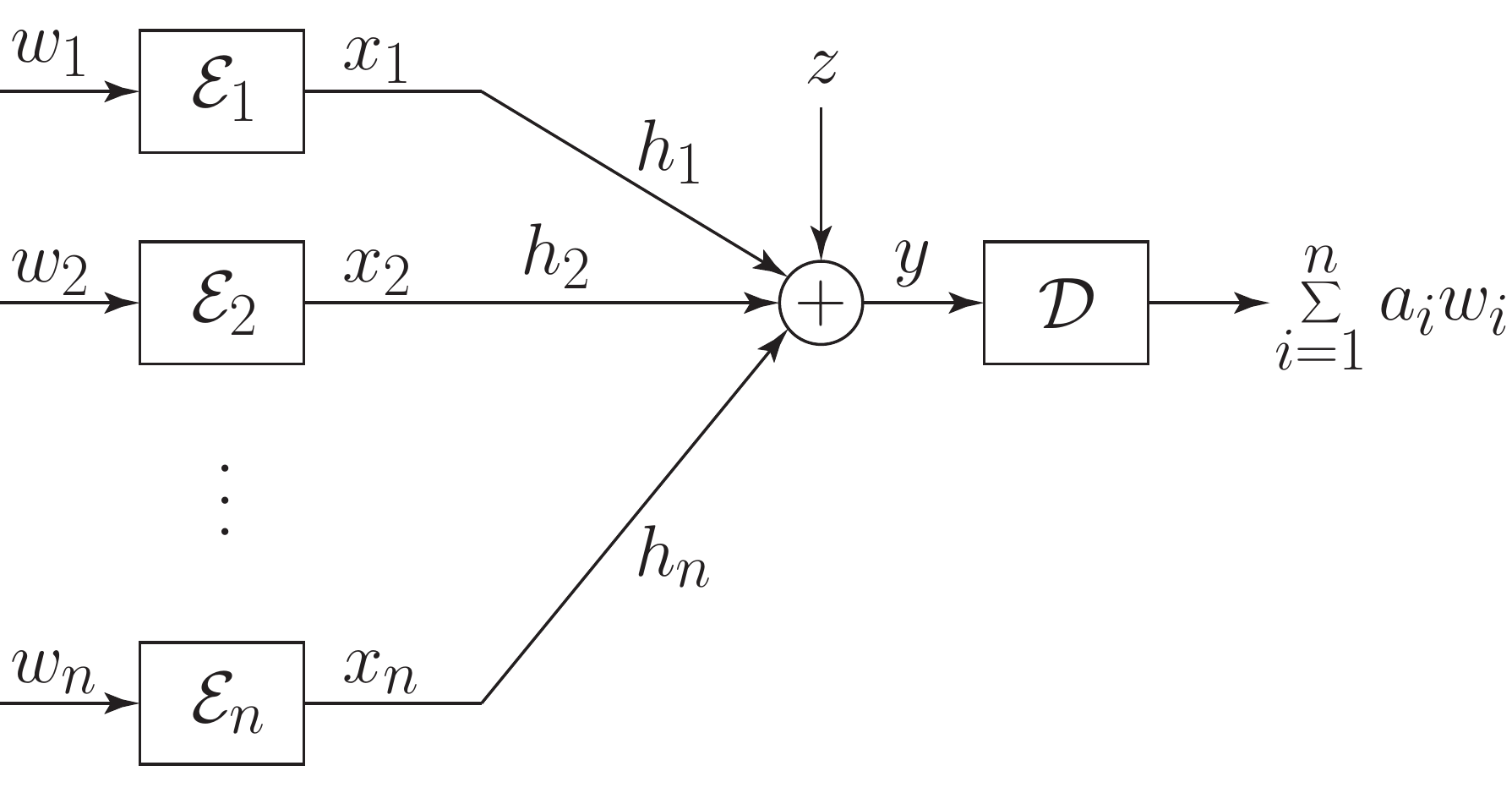}
\end{center}
\caption{$n$ transmitters send their messages to one relay. The relay decodes an integer linear combination of the codewords.}
\label{fig:fig2}
\end{figure}

Figure \ref{fig:fig2} demonstrates a Compute-and-Forward scenario, where ${n}$ transmitting nodes, each with transmission power $P$, share a wireless channel to send their messages to a relay node. We assume no knowledge of the channel states at the transmitters. The relay receives a noisy linear combination of the transmitted messages, namely

\begin{equation*}
{y} = \sum_{i=1}^nh_ix_i + z
\end{equation*}

where $x_i$ and $h_i$ respectively represent the signal transmitted by node $i$ and the effect of the channel from node $i$ to the decoder, and $z$ is the additive white Gaussian noise of unit variance. The relay then recovers $\sum a_i\omega_i$, an integer linear combination of the transmitted codewords.
It has been proved \cite{nazer2011compute} that the achievable rate of Compute-and-Forward satisfies: 
\begin{equation}
R({\bf h}, {\bf a}) =  \frac{1}{2}\log^+ \left(\left(\|{\bf a}\|^2 - \frac{P|{\bf h}^T{\bf a}|^2}{1+P\|{\bf h}\|^2} \right)^{-1} \right).
\label{eqn:CF}
\end{equation}

As evident from Equation \eqref{eqn:CF}, the achievable rate depends on the choice of the integer vector ${\bf a}$.
From the perspective of a single decoder, a reasonable choice for ${\bf a}$ is one that maximizes $R$: 
\begin{equation}
{\bf a}^* = \argmax_{{\bf a}\in \mathbb{Z}^n\backslash\{\bf 0\}}  \frac{1}{2}\log^+ \left(\left(\|{\bf a}\|^2 - \frac{P|{\bf h}^T{\bf a}|^2}{1+P\|{\bf h}\|^2} \right)^{-1} \right)
\label{eqn:CFopt}
\end{equation}
which can be simplified as
\begin{equation}
{\bf a}^* = \argmin_{{\bf a}\in \mathbb{Z}^n\backslash\{\bf 0\}}  f({\bf a})={\bf a}^T {\bf G a}
\label{eqn:CFmain}
\end{equation}
where
\begin{equation}
{\bf G}  =  {\bf I} - \frac{P}{1+P\|{\bf h}\|^2}{\bf h}{\bf h}^T
\label{eqn:gg}
\end{equation}
is a positive-definite matrix. Comparing this optimization problem with Equation \eqref{eqn:svp}, we see that Equation \eqref{eqn:CFmain} is an instance of the SVP. \\

\subsection{IP$^1$ Matrices and the Main Results}
The positive-definite matrix ${\bf G}$ in Equation \eqref{eqn:gg} falls in the following category of matrices which we refer to as IP$^1$. {This term is chosen to emphasize a decomposition of the form ${\bf I} - {\bf P}$ where ${\bf P}$ is of rank $1$.}
\begin{defn}
A positive-definite matrix ${\bf G}$ is called $\text{IP}^1$  if ${\bf G} ={\bf I} - \alpha{\bf v}{\bf v}^T$ where ${\bf v}$ is a normalized column vector in $\mathbb{R}^n$ and $0\le\alpha<1$ is a real number.
\label{def:ip1}
\end{defn}

The following theorem, albeit provable mostly by elementary manipulations of integer inequalities, establishes an important fact that provides the foundation of our SVP algorithm for IP$^1$ matrices. (The proof can be seen as a special case of Theorem \ref{thm1_dp1}. All the proofs are in Section \ref{sec:Appendices}.)

\begin{theorem}
{\label {THM1}}
Suppose ${\bf a}^*$ is the solution to \eqref{eqn:svp} for an IP$^1$ matrix ${\bf G}$. Then at least one of the following statements is true
\begin{itemize}
\item ${\bf a}^*$ satisfies
\begin{equation}
{\bf a^*} - \frac{1}{2}\mathbf{1}< {\bf v}x < {\bf a^*} + \frac{1}{2}\mathbf{1}
\label{eqn:inside_}
\end{equation}
and thus 
\begin{equation}
{\bf a^*} = \round{{\bf v}x}
\label{eqn:CFround_}
\end{equation}
for some $x \in \mathbb{R}^+$.
\item
{${\bf a^*}$ is a standard unit vector, up to a sign.}
\end{itemize}
\end{theorem}

It follows from Theorem \ref{THM1} that for the special lattices of interest, the shortest vector can be obtained by solving an optimization problem over only one variable. It is shown in \cite{nazer2011compute} that the solution to \eqref{eqn:CFmain} satisfies 
\begin{equation}
\|{\bf a^*}\| \le \sqrt{P\|{\bf h}\|^2+1}.
\label{eqn:CFbound}
\end{equation}

Equation (\ref{eqn:CFbound}) transforms into
\begin{equation}
\|{\bf a^*}\| \le \frac{1}{\sqrt{1-\alpha}}
\label{eqn:CFboundgeneral}
\end{equation}
for a general IP$^1$ matrix. Consequently, the search only has to be done over a bounded region {and in the vicinity of the line in the direction of ${\bf v}$ crossing the center. A separate examination of the standard unit vectors must also be performed.  This is a significant reduction in the number of candidate points compared to other exhaustive search algorithms that apply to general lattices as in \cite{hassibi2005sphere}, where such a structure is naturally absent and all the lattice points within an $n$-dimensional sphere must be enumerated. }

\begin{rem}
{\label{Remark1}}
The formula given by Theorem \ref{THM1} has some resemblance to the results of \cite{mckilliam2008linear} and \cite{mckilliam2010linear}. However the span of these works are Coxeter lattices \cite{martinet2013perfect} and the goal is to find faster algorithms for problems (CVP) which are already known to be polynomially solvable. {It is not difficult to see that the Gram matrix for Coxeter lattices have an IP$^k$ decomposition with small $k$ ($k\le 2$). Nevertheless, Coxeter  lattices are highly limited in structure. For instance, there are fewer than $n+1$ different Coxeter lattices in $\mathbb{R}^{n+1}$. All these lattices are rank-deficient and 
the solution to their SVP can be given in closed form. For instance, for $A^1_n$, the shortest vector is  $e_1 - \frac{1}{n+1}(\sum_{i=1}^{n+1}e_i)$ and for $A^{n+1}_n$, it is 
$e_1 - e_2$.}
\end{rem}

\subsection{SVP Algorithm for $\text{IP}^1$ Matrices}
In line with Theorem \ref{THM1} we define ${\bf a}({{x}}) = \round{{\bf v}x}$. Furthermore, let  $\psi = \frac{1}{\sqrt{1-\alpha}}$, so that we have $\|{\bf a^*}\| \le \psi$. Note that {Theorem \ref{THM1}} reduces the problem to a one-dimensional optimization task. Since every ${a}_i(x)$, the $i$'th element of the vector ${\bf a}({{x}})$, is a piecewise constant function of $x$, so is the objective function 
\begin{equation*}
f({\bf a}({x})) = \round{{\bf v}x}^T{\bf G}\round{{\bf v}x}.
\end{equation*}
Overall, the goal is to find a set of points which fully represent all the intervals in which $f(\cdot)$ is constant and choose the point that minimizes $f(\cdot)$. Being a piecewise constant function, $f(\cdot)$ can be represented as:
\begin{equation}
  f({\bf a}(x))=\begin{cases}
    r_i\;\; ,& \text{if $\xi_i< x< \xi_{i+1}$ ,  $i = 0,1,\dots$}\\
    s_i\;\; ,& \text{if $x = \xi_i$ ,  $i = 0,1,\dots$}\\
  \end{cases}
  \label{eqn:1}
\end{equation}
$\xi_i$'s are sorted real numbers denoting the points of discontinuity of $f(\cdot)$. Since $f(\cdot)$ is a continuous function of ${\bf a}$, these are in fact the discontinuity points of ${\bf a}(x)$ (or a subset of them) or equivalently the points where $a_i(x)$ is discontinuous, for some $i=1\dots n$. We can see from Equation \eqref{eqn:inside_} that any $x$ satisfying 
\begin{equation}
{a}_i^* - \frac{1}{2}< xv_i <{a}_i^* + \frac{1}{2}\;\;, \;\;i=1\dots n \;,\;v_i \neq 0
  \label{eqn:3}
\end{equation}
 minimizes $f(\cdot)$. As a result, $x$ belongs to the interior of an interval and not the boundary. Therefore, in the process of minimizing $f(\cdot)$, one can ignore the $s_i$ values in (\ref{eqn:1}), and find the minimizer of the objective function among the $r_i$ values.
\begin{equation}
\min_{{\bf a}\in \mathbb{Z}^n\backslash\{{\bf 0}\}} f({\bf a}) = \min_{i = 0,1\dots} r_i.
\end{equation}
Since $\frac{\xi_i+\xi_{i+1}}{2}$ belongs to the interior of the interval $(\xi_i , \xi_{i+1})$, we can rewrite $r_i$ as $r_i = f({\bf a}(\frac{\xi_i+\xi_{i+1}}{2}))$. Hence:
\begin{equation}
\min_{{\bf a}\in \mathbb{Z}^n\backslash\{{\bf 0}\}} f({\bf a}) = \min_{i = 0,1\dots} f({\bf a}(\frac{\xi_i+\xi_{i+1}}{2})).
  \label{eqn:4}
\end{equation}
As we discussed, $\xi_i$'s are the points where at least one of the elements of the vector ${\bf a}$ faces discontinuity. Since we have ${a_i}(x)= \round{{v}_ix}$, the discontinuity points of $a_i(x)$ are the points where $v_ix$ is a half-integer, or equivalently the points of the form $x= \frac{c}{|v_i|}$ where $c$ is a positive half-integer and $v_i\neq 0$. From Equation \eqref{eqn:CFboundgeneral} we can also see that $|a^*_i|\le \psi$ and therefore, $0<c\le\ceil{\psi}+\frac{1}{2}$. To conclude this argument, we write:

\begin{equation}
\xi_i \in \bigcup_{j=1}^n\Phi_j \;\; , \;\; i = 0,1,\cdots
\end{equation}
where 
\begin{eqnarray*}
\Phi_j &=&  \left\{ \frac{c}{|v_j|} \;\middle| \; 0<c \le\ceil{\psi}+\frac{1}{2}\;,\; c-\frac{1}{2} \in\mathbb{Z} \right\}\; ,\; v_j\neq 0,\\
\Phi_j &=& \font\msbm = msbm10 at 12pt
\hbox{\msbm \char 63} \; , \;v_j = 0\; , \; j= 1\dots n. 
\end{eqnarray*}

Thus, the algorithm starts by calculating the sets $\Phi_j$ and their union $\Phi$, sorting the elements of $\Phi$ and then running the optimization problem described by (\ref{eqn:4}). The standard unit vectors will also be individually checked. The number of elements in $\Phi_j$ is upper-bounded by $\ceil{\psi}+1$ and thus the number of elements in $\Phi$ is upper-bounded by $n(\ceil{\psi}+1)$. The value of $f(\cdot)$ can be calculated in constant time. This is thanks to the special structure of the matrix ${\bf G}$. In fact, the objective function can be rewritten as: 
\begin{equation}
f({\bf a}) = \sum{a_i^2} - \alpha\left(\sum a_iv_i\right)^2.
\label{eqn:simple}
\end{equation}
We keep track of every $a_i$ and the two terms $\sum{a_i^2}$ and $\sum a_iv_i$. Since the discontinuity points are sorted, at each step only one of the $a_i$'s changes and therefore the two terms can be updated in constant time. Consequently the new value of $f({\bf a})$ can also be calculated in constant time. (In order to remember which $a_i$ is being updated at each step, we assign a label to every member of ${\Phi}$ which indicates to which $\Phi_j$ it originally belonged).\\ 

\begin{algorithm}[H]
\caption[caption]{SVP for IP$^1$ matrices}
\begin{algorithmic}[1]
\Statex {{\bf Input:} The IP$^1$ matrix ${\bf G} = {\bf I} - \alpha {\bf v}{\bf v}^T.$}
 \Statex {\bf Output: }{$\bf {a^*}$ the solution to SVP for ${\bf G}$.}
\Statex ${\bf \underline {Initialization:}}$
 \State ${\bf u}_i\leftarrow$ standard unit vector in the direction of the $i$-th axis
 \State $\psi \leftarrow \frac{1}{\sqrt{1-\alpha}}$
\State $\Phi\leftarrow\font\msbm = msbm10 at 12pt
\hbox{\msbm \char 63}$
\State $f_{\min} \leftarrow \min(diag({\bf G}))$
\State ${\bf a^* }\leftarrow {\bf u}_{\argmin(diag({\bf G}))}$
\Statex{}
\Statex {\bf \underline {Phase 1:}}
\For{all $i \in \{1,\dots ,n\}$, and ${v_i \neq 0}$}
  \For{all ${c}$ , $0<{c} \le \ceil{\psi}+\frac{1}{2} \;,$  $ \;{c}-\frac{1}{2} \in\mathbb{Z} $}
\State	 ${x}\leftarrow\frac{c}{|v_i|}$
\State	 $\Phi \leftarrow \Phi \cup\{(x,i)\}$

\EndFor
\EndFor
\Statex{}
\Statex {\bf \underline {Phase 2:}}
\State sort $\Phi$ by the first element of the members (in an increasing order).
\State set $T_1 \leftarrow0$ , $T_2 \leftarrow 0$ and ${\bf a} \leftarrow {\bf 0}$.
\For{every $(x,j) \in\Phi$ (sweeping the set from left to right)}
\State	 $a_j \leftarrow a_j +\text{sign}(v_j)$
\State         $T_1 \leftarrow T_1 + 2a_j - 1$
\State         $T_2 \leftarrow T_2 + |v_j|$
\State         $f_{\text{new}} \leftarrow T_1 - \alpha T_2^2$
	\If {$f_{\text{new}}<f_{\min}$}
\State	 ${\bf a^* \leftarrow a}$
\State	 $f_{\min}\leftarrow f_{\text{new}}$

\EndIf
\EndFor
\State \Return {${\bf a^*}$}

\end{algorithmic}
\label{Algorithm1}
\end{algorithm}

It is easy to see that the complexity of the algorithm is determined by the sorting step. Since $\Phi$ has at most $n(\ceil{\psi}+1)$ members, the complexity is 
\begin{equation}
O(n\psi\log({n{\psi}}))
\label{eqn:complx}
\end{equation}
where $\psi = \frac{1}{\sqrt{1-\alpha}}$ for a general IP$^1$ matrix (Equation \eqref{eqn:CFboundgeneral}) and $\psi = \sqrt{1+P\|{\bf h}\|^2}$  for the Compute-and-Forward problem (Equation \eqref{eqn:CFbound}).\\
The procedure is summarized in Algorithm \ref{Algorithm1}. {To provide further insight, it is worth noting that the efficiency of the algorithm is due to two factors. Firstly, the number of candidate lattice points to be enumerated is bounded by a polynomial function of $n$. Secondly, thanks to the special structure of these lattice points (imposed by Theorem \ref{THM1}) their efficient enumeration is possible. To appreciate the importance of the second factor, an analogy can be made with sphere decoding \cite{ hassibi2005sphere} where the radius of the sphere is chosen in such a way that the number of lattice points within the sphere is polynomially bounded. However, performing an efficient exhaustive search over all the candidate points within the sphere remains as the main challenge. }

\begin{figure}
\begin{center}
\includegraphics[scale= 0.17]{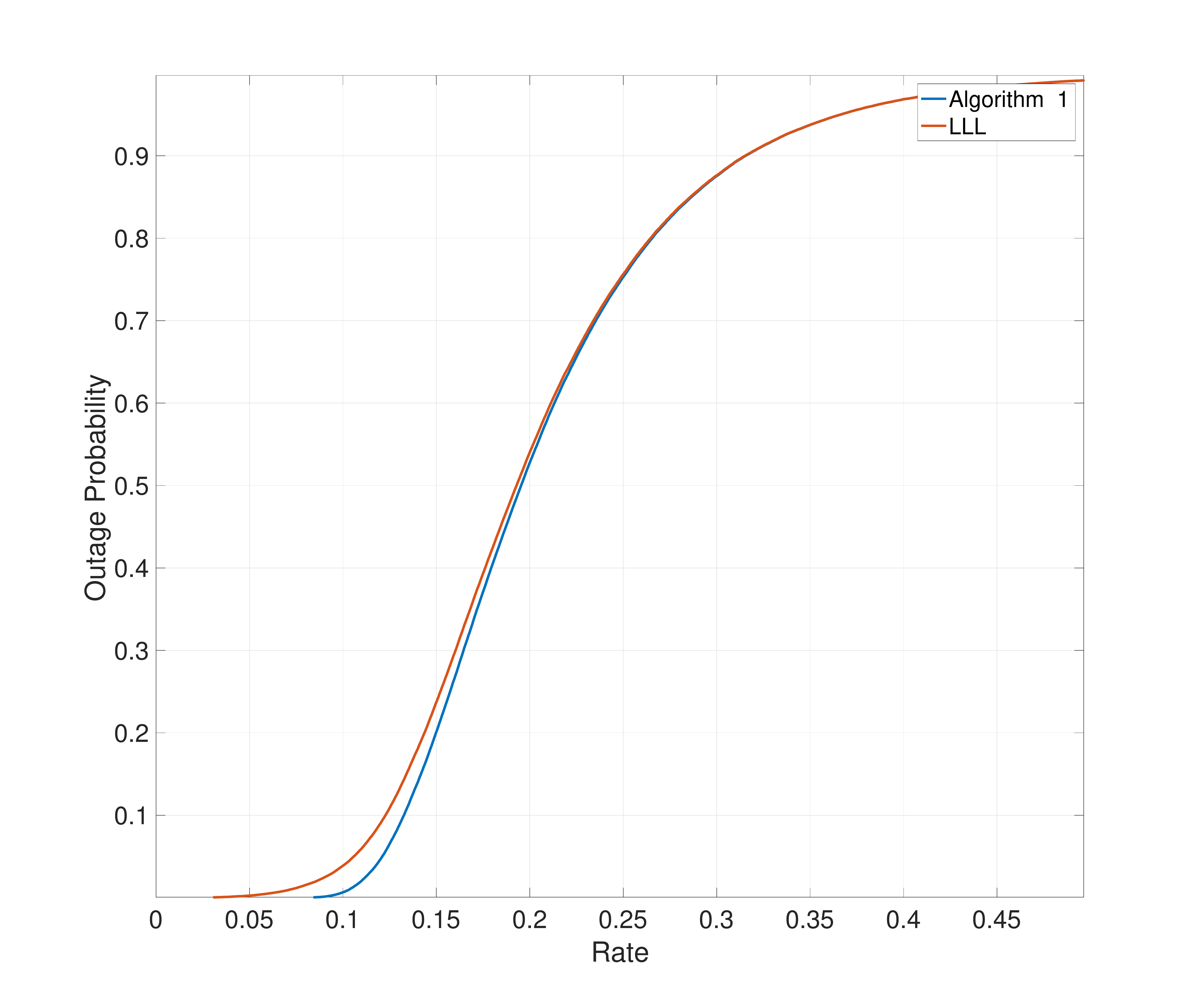}
\end{center}
\caption{A comparison of LLL vs. Algorithm \ref{Algorithm1} in terms of outage probability. Simulations are done with the following parameters:  $P=20$, $n = 20$ and ${\bf h}\sim{\cal N}({\bf 0},{\bf I})$. }
\label{fig:LLL}
\end{figure}

Prior to this work, it had been suggested  \cite{zhan2014integer}   to use approximation algorithms such as LLL in order to solve this instance of the SVP. It is well-known that despite its notorious worst-case performance (only exponential guarantee), LLL performs well in practice. In order to provide a quantitative judgment, we performed a MATLAB simulation by generating 100,000 realizations of channels distributed as ${{\bf h}\sim{\cal N}({\bf 0},{\bf I})}$ with $20$ users and with $P= 20$. Indeed, LLL returned the optimal solution in almost 91\% of the cases. Nevertheless, in the few cases where LLL fails, the difference in the achievable rate is rather significant (around 7\% on average). An outage rate curve has been provided in Figure \ref{fig:LLL} for the sake of comparison. In particular the gap in the outage probability is non-negligible at small transmission rates, i.e. $R \approx 0.1$ where the outage probability of Algorithm \ref{Algorithm1} is only around $0.6\%$ as opposed to LLL's $3.8\%$.  Many other algorithms exist in the literature as discussed in the introduction. Some of these works have excellent performance based on simulation results. Nevertheless a strict theoretical guarantee is missing on either their correctness or their complexity. (We refrain from presenting empirical complexity comparisons in terms of running time or otherwise owing to the strong dependence on the precise implementation of each algorithm.)

\subsection{Asymmetric Compute-and-Forward and DP$^1$ Matrices}
A slightly more general model compared to Definition \ref{def:ip1} is when the positive-definite matrix ${\bf G}$ is equal to 
\begin{equation}
{\bf G} = {\bf D} - \alpha{\bf v}{\bf v}^T
\label{eqn:dp1_dfn}
\end{equation}
where ${\bf D}$ is an arbitrary diagonal matrix with strictly positive diagonal elements, $\bf v$ is a normalized column vector and $0\le \alpha < 1$. We refer to such matrices as DP$^1$. Very similar to Theorem \ref{THM1} we have

\begin{theorem}
{\label {thm1_dp1}}
Suppose ${\bf a}^*$ is the solution to \eqref{eqn:svp} where ${\bf G}$ is DP$^1$, that is ${\bf G} = {\bf D} - \alpha{\bf v}{\bf v}^T$ as in Equation \eqref{eqn:dp1_dfn}. Then at least one of the following statements is true
\begin{itemize}
\item ${\bf a}^*$ satisfies
\begin{equation}
{\bf a^*} - \frac{1}{2}\mathbf{1}< {\bf D}^{-1}{\bf v}x < {\bf a^*} + \frac{1}{2}\mathbf{1}
\label{eqn:inside}
\end{equation}
and thus 
\begin{equation}
{\bf a^*} = \round{{\bf D}^{-1}{\bf v}x}
\end{equation}
for some $x \in \mathbb{R}^+$.
\item
{${\bf a^*}$ is a standard unit vector, up to a sign.}
\end{itemize}
Furthermore, 
\begin{equation*}
\|{\bf a}^*\|\le \psi =  \sqrt{\frac{G_{min}}{\lambda_{min}}}
\end{equation*}
where $G_{min}$ is the smallest diagonal element of the matrix ${\bf G}$ and $\lambda_{min}$ is the smallest eigenvalue of ${\bf G}$.
\end{theorem}

Algorithm \ref{Algorithm1} can then be readily extended to solve the SVP for DP$^1$ matrices. The following modifications are necessary. In step 8, $v_i$ should be replaced by $\frac{v_i}{D_{ii}}$. Furthermore, $f({\bf a})$ is no longer of the form \eqref{eqn:simple}, but instead
\begin{equation}
f({\bf a}) = \sum{D_{ii}a_i^2} - \alpha\left(\sum a_iv_i\right)^2.
\end{equation}
Therefore, at step 16, $T_1$ should be updated as $T_1 = T_1 +D_{jj}(2a_j -1)$. Finally, the value of $\psi$ should be replaced with the more general expression
$\psi =  \sqrt{\frac{G_{min}}{\lambda_{min}}}$ to represent the new bound on the value of $\|{\bf a}^*\|\le \psi.$ With this change in value of $\psi$, the complexity again follows \eqref{eqn:complx}.

The interest in DP$^1$ matrices originates from their application in Asymmetric Compute-and-Forward \cite{zhu2014asymmetric}. Concisely, if in the Compute-and-Forward scheme we allow the transmitters to transmit at different rates, the achievable computation rate of the $k$'th transmitter is proved  \cite{zhu2014asymmetric} to be equal to:
\small\begin{equation}
R_k({\bf h},{\bf a},{\bf B})  = \left[\frac{1}{2}\log\left(\|{\bf Ba}\|^2-\frac{P|{\bf h}^T{\bf Ba}|^2}{1+P\|{\bf h}\|^2}\right)^{-1}+\frac{1}{2}\log B_{kk}^2\right]^+
\end{equation}\normalsize
where ${\bf B}$ is an arbitrary diagonal matrix with positive diagonal elements. These diagonal elements are chosen by the respective transmitters based on their channel state information. Clearly, the integer vector ${\bf a}$ that maximizes the achievable rate is the same for all of the transmitters: 
\begin{equation}
{\bf a}^* = \argmin_{{\bf a}\in \mathbb{Z}^n\backslash\{{\bf 0}\}}{\bf a}^T{\bf Ga},
\end{equation}
where ${\bf G}$ is given by:
\begin{equation}
{\bf G} = {\bf B}\left({\bf I} - \frac{P}{1+P\|{\bf h}\|^2}{\bf h}{\bf h}^T\right){\bf B} 
\end{equation}
which is a DP$^1$ matrix. Therefore, the extension of  Algorithm \ref{Algorithm1} can be used to find the vector ${\bf a}$ which simultaneously maximizes the achievable rate for all transmitters. 

It should be noted that for DP$^1$ matrices, the algorithm requires the decomposition of ${\bf G}$ as ${\bf D} - \alpha{\bf v}{\bf v}^T$. This information might be given a priori as with Asymmetric Compute-and-Forward or the decomposition could be found using the so called diagonal and low rank matrix decomposition techniques studied in \cite{saunderson2012diagonal,shapiro1982weighted}.
 
\section{$\text{IP}^{k}$ Matrices and Integer-Forcing}
\label{sec:IPK}
\subsection{Integer-Forcing}
In this section we provide a generalization of Theorem \ref{THM1} and the corresponding algorithm by relaxing several constraints that we imposed on the structure of the Gram matrix ${\bf G}$. The generalized theorem can be applied to maximize the achievable computation rate of Integer-Forcing studied in \cite{zhan2014integer}. The scenario is very similar to the previous section, with the difference that the relay node now has multiple antennas. Our objective remains the same: decode the best integer linear combination of the received codewords. Assume there are $n$ transmitters with transmission power $P$ and the receiver node has $k$ antennas. Let ${\bf h}_i$ be the channel vector from the transmitting nodes to the $i$-th antenna of the relay. Also, let ${\bf H}$ be the $n\times k$ matrix whose columns are the ${\bf h}_i$ vectors. It directly follows from the results of \cite{zhan2014integer} that the achievable computation rate satisfies the following equation:
\begin{equation}
R({\bf a}) = - \frac{1}{2}\log {\bf a}^T{\bf Ga}
\end{equation}
where
\begin{equation}
{\bf G} = {\bf WRW}^T.
\label{eqn:CFMIMO}
\end{equation}
Here ${\bf W \;\;}$ is a unitary matrix in $\mathbb{R}^{n\times n}$ whose columns are the eigenvectors of ${\bf HH}^T$, and ${\bf R}$ is a diagonal square matrix with the first $k$ diagonal elements satisfying
$$r_i = \frac{1}{1+P\gamma_i^2}\;\;,\;\; i=1\cdots k$$
 and the last $n-k$ diagonal elements equal to 1. Finally, $\gamma_i^2$ is the $i$-th eigenvalue of ${\bf HH}^T$ (same order as the columns of ${\bf W}$).\\
 Our goal is to find 
 $${\bf a}^* = \argmin_{{\bf a}\in \mathbb{Z}^n\backslash\{\bf 0\}}  {\bf a}^T {\bf G a}$$
as in the single antenna case.
\subsection{IP$^k$ Matrices and the Main Results}
We first mention a generalization of Theorem \ref{THM1} and next we show that the Gram matrix which appears in Equation \eqref{eqn:CFMIMO} satisfies the constrains of the new theorem. 
To begin with, we define the following:

\begin{defn}[$\text{IP}^k$ matrices]
A positive-definite matrix ${\bf G}$ is called $\text{IP}^k$  if ${\bf G =I - P}$ where ${\bf P}$ is a positive semi-definite matrix of rank $k$ and ${\bf I}$ is the identity matrix.\\
\label{def:ipk}
\end{defn}
We find it convenient to write ${\bf G}$ as 
\begin{equation}
{\bf G} = {\bf I} - {\bf VV}^T
\label{eqn:ultdecom}
\end{equation}
 where ${\bf V}$ is an $n\times k$ matrix. Such a decomposition is not unique, but our arguments will be valid regardless of how the matrix ${\bf V}$ is chosen.
 
 Theorem \ref{THM2} is a generalization of Theorem \ref{THM1} with a similar claim: the SVP for IP$^k$ matrices can be reduced to a search problem over only $k$ dimensions. 

\begin{theorem}
{\label {THM2}}
Suppose ${\bf a}^*$ is the solution to \eqref{eqn:svp} where ${\bf G}$ is IP$^k$, that is ${\bf G} = {\bf I} - {\bf VV}^T$ as in Equation \eqref{eqn:ultdecom}. Then at least one of the following statements is true
\begin{itemize}
\item There exists a vector ${\bf x}\in \mathbb{R}^{k}$ such that  ${\bf a}^* - \frac{1}{2}\mathbf{1}< {\bf Vx} < {\bf a}^* + \frac{1}{2}\mathbf{1}$ and thus ${\bf a}^* = \round{{\bf Vx}}$. 
\item ${\bf a}^*$ is a standard unit vector, up to a sign.
\end{itemize}
Furthermore, $\|{\bf a}^*\| \le \psi = \sqrt{\frac{G_{min}}{\lambda_{min}}}$ where $G_{min}$ is the smallest diagonal element of ${\bf G}$ and $\lambda_{min}$ is the smallest eigenvalue of ${\bf G}$.
\end{theorem}

Note that Theorem \ref{THM1} is a special case of Theorem \ref{THM2} where $k=1$. The bound on the norm of ${\bf a}^*$ turns into $\frac{1}{\sqrt{1-\alpha}}$, for IP$^1$ matrices since we have $\lambda_{min} = 1-\alpha$ and $G_{min} \le 1$.

The Gram matrix in equation (\ref{eqn:CFMIMO}) also satisfies the constraints of Theorem \ref{THM2}: Since ${\bf W}$ is a unitary matrix, ${\bf G}$ can be rewritten as ${\bf I} -  {\bf W(I-R)W}^T$. The matrix ${\bf W(I-R)W}^T$ is of rank $k$ (since ${\bf I - R}$ has only $k$ non-zero diagonal entries), and positive semi-definite. The bound given by the theorem translates into $\|{\bf a}^*\| \le \sqrt{1 + P\gamma_{max}^2}$ where $\gamma_{max}$ is the maximum $\gamma_i$ value. This is because $G_{min}\le 1$ and the eigenvalues of ${\bf G}$ are equal to $\frac{1}{1 + P\gamma_i^2}$ (with the same eigenvectors as ${\bf HH}^T$) or 1.

We will now show how to solve the SVP for an IP$^k$ matrix using Theorem \ref{THM2}.
\subsection{SVP Algorithm for $\text{IP}^k$ Matrices}
\label{sec:alg2}
Similar to the case $k=1$ we see that
\begin{equation*}
f({\bf a}({\bf x})) = \round{{\bf Vx}}^T{\bf G}\round{{\bf Vx}}
\end{equation*}
is piecewise constant as a function of the vector ${\bf x}$ (this is because $\round{{\bf Vx}}$ is a piecewise constant function of ${\bf x}$). Our objective is very similar to before: enumerate all the regions in space in which the objective function $f(\cdot)$ is constant and choose the one that minimizes $f(\cdot)$. From Theorem \ref{THM2} we know that the vector ${\bf a}^*$ satisfies the $2n$ inequalities:
$${\bf a^*} - \frac{1}{2}\mathbf{1}< {\bf Vx} < {\bf a^*} + \frac{1}{2}\mathbf{1}$$
for some ${\bf x}\in \mathbb{R}^k$. In other words, ${\bf x}$ belongs to the interior of the cell described by these half-spaces. By analogy to the case $k=1$ we aim at enumerating all such cells and finding the one which minimizes the objective function. To start with, we observe that each such cell is bounded by a set of hyperplanes of the form: 
$${\bf V}_{\{i\}}{\bf x} = c$$
where $c$ is a half integer. Due to the bound given by Theorem \ref{THM2} we could show that the hyperplanes are restricted to $|c|\le (\ceil{\psi}+\frac{1}{2})$ which gives us a total of $n(2\ceil{\psi}+2)$ hyperplanes. 
The problem of efficient enumeration of all the cells resulting from a partitioning of the space by a set of hyperplanes is the subject of a field called Hyperplane Arrangements. Very efficient algorithms have been developed over the past few decades. The general idea behind most of these algorithms is the following: we assign a normal vector with a specific direction to every hyperplane. Since a cell is bounded by hyperplanes, it must be entirely located on one side of each hyperplane. Therefore, a cell can be represented by a sign vector ${\bf \nu}$ of length $m$ where $m$ is the number of hyperplanes. Each $\nu_i$ is either $+1$ or $-1$ depending on whether the cell is located on the positive side or the negative side of the corresponding hyperplane. Although there are $2^m$ possible configurations for the sign vector ${\bf \nu}$, at most $O(m^k)$ cells are created by the intersection of $m$ hyperplanes. The enumeration algorithm will aim at finding those sign vectors that correspond to the actual cells. We will discuss two existing algorithms due to \cite{gerstneralgorithms} and \cite{sleumer1999output}. The first one is very simple to understand and implement but it might face numerical issues in case of degeneracies, i.e. when there are more than $k+1$ hyperplanes intersecting at the same point. It is clear however that in practice, such an event occurs with probability zero. The second algorithm is slightly more complicated but it covers degeneracies too.   

\subsubsection{Simple Cell Enumeration Algorithm \cite{gerstneralgorithms}}
The idea is to first find all the vertices of the cells by finding the intersection of every $k$ hyperplanes. We can represent each vertex by a sign vector of length $m$ where each entry belongs to $\{-1,0,1\}$ depending on whether the vertex is on the left hand side of the corresponding hyperplane, located on it, or on its right hand side.  Assuming there are no degeneracies, each such vertex belongs to exactly $k$ hyperplanes (and not more). Consequently, every vertex has a sign vector with exactly $k$ elements equal to zero. In this case, each vertex belongs to exactly $2^k$ cells whose sign vectors can be found by taking all possible assignments of $\{-1,1\}$ to the zero elements of the sign vector of the vertex. Repeating this procedure for every vertex, we will have enumerated every cell of the arrangement, albeit in a redundant way. The running time of the algorithm is $O(m^{k+1})$. In case of degeneracies, \cite{gerstneralgorithms} suggests that we slightly perturb the hyperplanes which however may not perform very well for highly degenerate matrices due to numerical issues. 
\subsubsection{Output-sensitive Cell Enumeration \cite{sleumer1999output}}
We represent every cell by a node in a graph. Two nodes are connected by an edge if and only if the corresponding cells are adjacent in space; in other words, if the sign vectors of the two cells differ in exactly one element. Intuitively such a graph is always connected. The algorithm aims at finding a spanning tree of this graph rooted at an arbitrary node. It also provides an interior point of each cell (here we are only interested in these interior points and not the sign vectors). There are two challenges. Firstly, we do not have a global knowledge of the graph and starting from each node we need to discover the neighboring nodes in an efficient way. Secondly, in order to form the spanning tree we must uniquely determine the parent of each node. The Output-sensitive Cell Enumeration algorithm in \cite{sleumer1999output} uses two subroutines, namely adjlist() and parent() to address these two problems. The pseudocode is provided in Algorithm \ref{hyperplane}.  

The overall complexity of the algorithm is $O(m|C|)$ where $m$ is the number of hyperplanes and $C$ is the number of cells which in turn is upper-bounded by $O(m^k)$. In our case, the number of hyperplanes is $n(2\ceil{\psi}+2)$. Therefore the algorithm runs in $O\left(n^{k+1}(2\ceil{\psi}+2)^{k+1}\right)$.

\begin{algorithm}[H]
\begin{algorithmic}[1]
\Statex {{\bf Input:} The root cell ${\bf c}$ represented by its sign vector. The hyperplanes given by the matrix ${\bf A}$ and the vector ${\bf b}$}
\Statex {{\bf Output:} An interior point of each cell in the subtree rooted at ${\bf c}$.}
\Statex{}
\Statex {\bf \underline {begin}}
\State Output an interior point of ${\bf c}$.
\State Find adjlist(${\bf c}$); the list of all neighbors of ${\bf c}$
\For {each ${\bf d}\in \text{adjlist}({\bf c})$}
\If {parent(${\bf d}$) = ${\bf c}$}
 \State CellEnum(${\bf A},{\bf b},{\bf d}$)
\EndIf
\EndFor
\end{algorithmic}
\caption[caption]{CellEnum(${\bf A},{\bf b},{\bf c}$)}
\label{hyperplane}
\end{algorithm}

\begin{algorithm}[H]
\begin{algorithmic}[1]
\Statex {{\bf Input:} The $\text{IP}^k$ Matrix ${\bf G} = {\bf I} - {\bf VV}^T$.}
\Statex {{\bf Output:} ${\bf a}^*$ the solution to the SVP for $\bf G$. }
\Statex{}
\Statex ${\bf \underline {Initialization}}$
 \State ${\bf u}_i:=$ standard unit vector in the direction of $i$-th axis
 \State $\psi := \sqrt{\frac{G_{\min}}{\lambda_{min}}}$
\State $f({\bf a}):={\bf a}^T{\bf G}{\bf a}$
\State $f_{\min} = \min(diag({\bf G}))$
\State ${\bf a^* }= {\bf u}_{\argmin(diag({\bf G}))}$
\Statex{}
\Statex {\bf \underline {begin}}
\State Form the matrix $\bar{\bf V}=
\left[
\begin{array}{c|c|c}
{\bf V}^T & \cdots & {\bf V}^T\\
\end{array}
\right]^T$ by repeating $\bf V$, $(\ceil{\psi}+1)$ times. Then $\bar{\bf V}\leftarrow
\left[
\begin{array}{c|c}
\bar{\bf V}^T & -\bar{\bf V}^T\\
\end{array}
\right]^T$.
\State Form the vector $\bar{\bf c}=
\left[
\begin{array}{c|c|c}
{\bf c}_{1}^T & \cdots &{\bf c}_{L}^T\\
\end{array}
\right]^T$ where $L = \ceil{\psi}+ 1$  and ${\bf c}_i = (\frac{1}{2} - i)\mathbf{1}$ and $\mathbf{1}$ is of length $n$. Then $\bar{\bf c}\leftarrow
\left[
\begin{array}{c|c}
\bar{\bf c}^T & \bar{\bf c}^T\\
\end{array}
\right]^T$.
\State $\Phi =$ CellEnum($\bar{\bf V},\bar{\bf c},\{+,...,+\}$)
\For {each ${\bf d}\in \Phi$}
\State Find ${\bf a} = \round{\bf Vd}$
\If {$f({\bf a})< f_{\text{min}}$ AND $\bf a$ is not the all-zero vector}
\State Set ${\bf a}^* = {\bf a}$.
\State Set $f_{\text{min}} = f({\bf a})$. 
\EndIf
\EndFor

\end{algorithmic}
\caption[caption]{SVP for IP$^k$ matrices}
\label{Algorithm2}
\end{algorithm}

In order to ensure that all the cells are enumerated, it is necessary to make the first call to Algorithm  \ref{hyperplane} with the parameter ${\bf c} = \{+,\dots,+\}$, that is the all plus sign vector. To guarantee that the all plus sign vector corresponds to an actual region, we change the direction of the hyperplanes in such a way that the origin is on the plus side of every hyperplane.
Algorithm \ref{Algorithm2}   will find the solution to the SVP for an IP$^k$ matrix by first finding a list of all the hyperplanes, then calling Algorithm \ref{hyperplane}   in order to find an interior point of each cell, and finally calculating the value of the objective function over an interior point of each such cell in $O(kn)$. Since there are at most $O(m^k)$ cells, the complexity of Algorithm \ref{Algorithm2} is the same as Algorithm \ref{hyperplane} (for constant $k$) that is 
\begin{equation}
O\left(n^{k+1}(2\ceil{\psi}+2)^{k+1}\right).
\label{eqn:complexity}
\end{equation}

\subsection{Asymmetric MIMO Compute-and-Forward and DP$^k$ Matrices}
Similar to IP$^1$ matrices, we can slightly extend the results by replacing the identity matrix in Equation \eqref{eqn:ultdecom} with an arbitrary diagonal matrix with strictly positive diagonal values. More precisely, we define the DP$^k$ matrices as positive-definite matrices of the form 
\begin{equation}
{\bf G} = {\bf D} - {\bf VV}^T
\label{eqn:dpk_dfn}
\end{equation}

where ${\bf V}\in \mathbb{R}^{n\times k}$ and ${\bf D}$ is diagonal with strictly positive diagonal elements. The following Theorem holds:

\begin{theorem}
{\label {thm_dpk}}

Suppose ${\bf a}^*$ is the solution to \eqref{eqn:svp} where ${\bf G}$ is DP$^k$, that is ${\bf G} = {\bf D} - {\bf VV}^T$ as in Equation \eqref{eqn:dpk_dfn}. Then at least one of the following statements is true
\begin{itemize}
\item There exists a vector ${\bf x}\in \mathbb{R}^{k}$ such that  ${\bf a}^* - \frac{1}{2}\mathbf{1}< {\bf D}^{-1}{\bf Vx} < {\bf a}^* + \frac{1}{2}\mathbf{1}$ and thus ${\bf a}^* = \round{{\bf D}^{-1}{\bf Vx}}$. 
\item ${\bf a}^*$ is a standard unit vector, up to a sign.
\end{itemize}
Furthermore, $\|{\bf a}^*\| \le \psi = \sqrt{\frac{G_{min}}{\lambda_{min}}}$ where $G_{min}$ is the smallest diagonal element of ${\bf G}$ and $\lambda_{min}$ is the smallest eigenvalue of ${\bf G}$.
\end{theorem}

Algorithm \ref{Algorithm2} can be reused with the following modifications in order to solve the SVP for DP$^k$ matrices. In step 6 we now have $\bar{\bf V}$ as the vertical concatenation of the matrix ${\bf D}^{-1}{\bf V}$. Moreover, step 10 should be replaced by ${\bf a } = \round{{\bf D}^{-1}{\bf Vd}}$. The complexity of the algorithm is again given by \eqref{eqn:complexity}.

This new version of the algorithm can help us with maximizing the achievable computation rate of all transmitters in an asymmetric MIMO Compute-and-Forward scheme \cite{zhu2014asymmetric}, assuming that the receiver aims at decoding a single integer linear combination of transmitted messages. In this case the achievable computation rate for the $k$'th transmitter is given \cite{zhu2014asymmetric} by
\begin{equation}
R_k({\bf h},{\bf a},{\bf B})  =-\frac{1}{2}\log\frac{{\bf a}^T{\bf BWRW}^T{\bf Ba}}{B_{ii}^2}
\end{equation}
where ${\bf W}$ and ${\bf R}$ are as in \eqref{eqn:CFMIMO} and ${\bf B}$ is an arbitrary diagonal matrix with positive diagonal elements selected at the transmitters. We can simultaneously maximize the achievable rate for all transmitters by solving the SVP for the matrix ${\bf G} ={\bf BWRW}^T{\bf B}$. We know from earlier discussion that ${\bf WRW}$ is IP$^k$ from which it directly follows that ${\bf BWRW}^T{\bf B}$ is DP$^k$. Hence, our modified algorithm can be used to solve this instance of SVP. 
\section{Approximate SVP and CVP for $\widetilde{\text{IP}}_\gamma^{k}$ Matrices}
\label{sec:almostIPK}
\subsection{$\widetilde{\text{IP}}^k_\gamma$ Matrices}

In this chapter we introduce a larger class of lattices for which both the SVP and CVP can be approximated up to a constant factor. As evident from Definition \ref{def:ipk}, the eigenvalues of an IP$^k$ matrix have a very particular structure: $n-k$ eigenvalues are equal to $1$ and the remaining $k$ eigenvalues are between $0$ and $1$. The main idea here is to relax this rather tight constraint and allow the eigenvalues to change within a neighborhood of these values. We will show that if these variations are small, the solution to the SVP and the CVP can be approximated within a constant factor. We start by defining the concept of IP$^k$-approximation of positive-definite matrices.

\begin{figure}[h]
\begin{center}
\includegraphics[width=6cm]{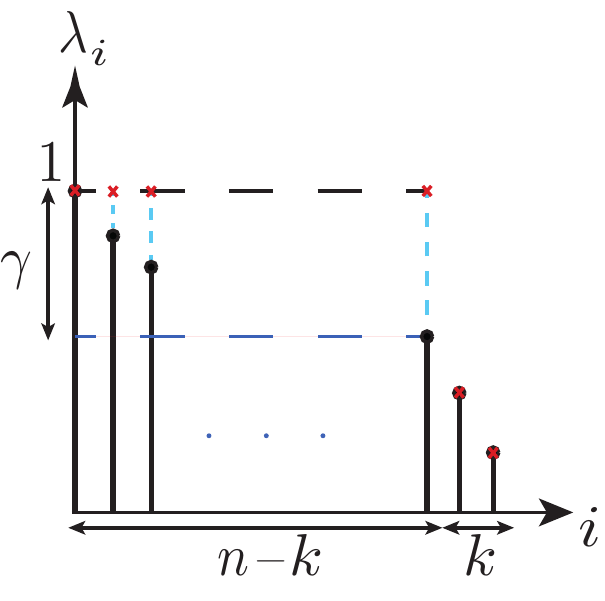}
\end{center}
\caption{Eigenvalues of an $\widetilde{\text{IP}}^k_\gamma$ matrix (black circles) and its $\tilde{\text{IP}}^k$-approximation (red crosses).}
\label{fig:fig1}
\end{figure}

\begin{defn}
Let ${\bf Q}\in \mathbb{R}^{n\times n}$ be a symmetric matrix with eigenvalues in $(0,1]$ and with the following eigendecomposition:
$${\bf Q = V}^T{\bf \Lambda V}.$$
For $k=0,\dots ,n$, we define the $\text{IP}^k$-approximation of ${\bf Q}$ as:
$$\mathcal{I}_k({\bf Q})={\bf V}^T\hat{\bf \Lambda}{\bf V}$$
where $\hat{\bf \Lambda}$ is obtained by setting the largest $n-k$ diagonal elements of $\bf \Lambda$ to one.
\end{defn}
The assumption that the eigenvalues must be less than or equal to one is not of fundamental importance. If the eigenvalues of a Gram matrix ${\bf G}$ are larger than one, we can normalize all the eigenvalues by the largest one. This translation does not have any effect on the solution of SVP (for CVP we will also have to scale the vector ${\bf y}$).
\begin{defn}
A symmetric matrix ${\tilde{\bf G}}\in \mathbb{R}^{n\times n}$ with eigenvalues in $(0,1]$ is called $\widetilde{\text{IP}}^k_\gamma$ if $\mathcal{I}_k({\tilde{\bf G}})-\tilde{\bf G}$ has all its eigenvalues smaller or equal to $\gamma$, where $\gamma$ is a constant satisfying $0\le\gamma<1$.
\end{defn}
In particular, the largest $n-k$ eigenvalues of an $\widetilde{\text{IP}}^k_\gamma$ matrix cannot be arbitrarily close to zero. They must be within a constant ($\gamma$) gap of one.
Figure \ref{fig:fig1} represents the sorted eigenvalues of an $\widetilde{\text{IP}}^k_\gamma$ matrix (marked by black circles) and its $\text{IP}^k$-approximation (red crosses).
\subsection{Approximate SVP Algorithm for $\widetilde{\text{IP}}^k_\gamma$ Matrices}
The following Theorem establishes a close connection between the solution of the SVP for an $\widetilde{\text{IP}}^k_\gamma$ matrix and for its $\text{IP}^k$-approximation. 
\begin{theorem}
Let 
$$f({\bf a}) = {\bf a}^T{\tilde{\bf G}}{\bf a}$$
where ${\tilde{\bf G}}$ is $\widetilde{\text{IP}}^k_\gamma$. Assume ${\bf a}^*$ is the solution to 
\begin{equation*}
{\bf a}^* = \argmin_{{\bf a\in Z}^n\backslash \{\bf 0\}}f({\bf a})
\end{equation*}
and $\hat{\bf a}$ satisfies
\begin{equation*}
\hat{{\bf a}} = \argmin_{{\bf  a\in Z}^n\backslash \{\bf 0\}} {\bf a}^T\mathcal{I}_k(\tilde{\bf G}){\bf a}
\end{equation*}
Then we have: 
\begin{equation}
f(\hat{\bf a}) < \frac{1}{1-\gamma}f({\bf a}^*).
\end{equation}
\label{thm:mainthm}
\end{theorem}
Theorem \ref{thm:mainthm} suggests that instead of solving the SVP for the $\widetilde{\text{IP}}^k_\gamma$ matrix ${\tilde{\bf G}}$, we can solve the problem for $\mathcal{I}_k(\tilde{\bf G})$, the ${\text{IP}}^k$-approximation of ${\tilde{\bf G}}$. The solution achieves a constant ($\frac{1}{1-\gamma}$) approximation factor on the original problem. As we saw in Section \ref{sec:IPK} the SVP for $\mathcal{I}_k(\tilde{\bf G})$ (which is an IP$^k$ matrix) can be found in polynomial complexity. Therefore, Algorithm \ref{approxSVP} is proposed to approximate the SVP for $\tilde{\bf G}$. A trivial improvement here would be to perform this minimization task over the original objective function. In other words, we can change line $3$ of Algorithm $\ref{Algorithm2}$ to $f({\bf a}):={\bf a}^T\tilde{\bf G}{\bf a}$.
\begin{algorithm}[H]
\caption[caption]{$\frac{1}{1-\gamma}$-approximation algorithm for the SVP for $\widetilde{\text{IP}}^k_\gamma$ matrices}

\begin{algorithmic}[1]
\Statex {\bf Input:} $\widetilde{\text{IP}}^k_\gamma$ Matrix ${\tilde{\bf G}}$
 \State Find $\hat{\bf G} = \mathcal{I}_k(\tilde{\bf G})$
 \State \Return {$\hat{\bf a}$ , the output of Algorithm \ref{Algorithm2} applied on $\hat{\bf G}$}. 
\end{algorithmic}
\label{approxSVP}
\end{algorithm}

Finding the IP$^k$ approximation of a matrix can be done in the same complexity order as finding its eigen-decomposition, that is $O(n^3)$, therefore, the overall complexity still follows Equation \eqref{eqn:complexity} (for $k>1$)
where $\psi$ is equal to $\sqrt{\frac{G_{min}}{\lambda_{min}}}$, the ratio of the smallest diagonal element and the smallest eigenvalue of the matrix $\hat{\bf G}$. Although ${\lambda_{min}}$ is equal for the two matrices $\hat{\bf G}$ and $\tilde{\bf G}$, the parameter $G_{min}$ is in general larger for the matrix $\hat{\bf G}$ compared to $\tilde{\bf G}$. Nonetheless, it is evident that all the diagonal entries of $\hat{\bf G}$ are upper bounded by $1$. Therefore we could still claim that as long as  $\frac{1}{\lambda_{min}}$ is upper bounded by a polynomial function of $n$, the algorithm runs in polynomial complexity.
\subsection{Extension of the Results to the CVP}
Our results can be readily generalized to the CVP. For instance, the CVP for an $\text{IP}^k$ matrix can be solved in an almost identical approach to the SVP. Furthermore, a similar constant-factor approximation for the CVP for $\widetilde{\text{IP}}^k_\gamma$ matrices can be obtained. This is particularly interesting since in general, the algorithms that are used for solving the CVP are more sophisticated compared to the SVP.

The following theorem tells us that the same dimensionality reduction that appears in the SVP for $\text{IP}^k$, also holds for the CVP:

\begin{theorem}
Suppose ${\bf A}\in \mathbb{R}^{n\times n}$ and ${\bf G =A}^T{\bf A }$ is $\text{IP}^k$, that is 
$${\bf G} = {\bf I - VV}^T$$
as in Equation \eqref{eqn:ultdecom}.
The solution to the Closest Vector Problem
\begin{equation*}
{\bf a}^* = \argmin_{{\bf a}\in \mathbb{Z}^n} \|{\bf Aa-y}\|^2
\end{equation*}
satisfies:
\begin{equation}
{\bf a}^* = \left\lceil{{\bf V}{{\bf x + A}^T{\bf y}}}\right\rfloor
\label{eqn:round}
\end{equation}
for some ${\bf x}\in \mathbb{R}^k$. Furthermore, 
\begin{equation}
\|{\bf a}^* - {\bf A}^{-1}{\bf y}\| \le \psi = \sqrt{\frac{{G}_{max}}{\lambda_{min}}}
\label{eqn:bound}
\end{equation}
where ${G}_{max}$ and $\lambda_{min}$ are the largest diagonal element and the smallest eigenvalue of the matrix ${\bf G}$, respectively.
\label{thm:cvpthm}
\end{theorem}

The same Hyperplane Arrangement technique as in Section \ref{sec:IPK} can be applied here. The main difference is that now the hyperplanes are shifted compared to the case of SVP. Similar to Algorithm \ref{Algorithm2} we need to ensure that the all plus sign vector corresponds to an actual region. We will do this by changing the direction of the hyperplanes in such a way that the origin is on the plus side of every hyperplane.
See Algorithm \ref{cvpExact}.
\begin{algorithm}[H]
\begin{algorithmic}[1]
\Statex {{\bf Input:} The $\text{IP}^k$ Matrix ${\bf G} = {\bf I} - {\bf VV}^T$ and the latice basis ${\bf A}$ that satisfies ${\bf G}= {\bf A}^T{\bf A}$ and the vector ${\bf y}\in\mathbb{R}^n$.}
\Statex {{\bf Output:} ${\bf a}^*$ the solution to the CVP for $\bf A$ and ${\bf y}$. }
\Statex{}
\Statex ${\bf \underline {Initialization}}$
 \State $\psi := \sqrt{\frac{G_{max}}{\lambda_{min}}}$
\State $f({\bf a}):={\bf a}^T{\bf G}{\bf a} -2 {\bf y}^T{\bf Aa} + {\bf y}^T{\bf y}$
\State $f_{\min} = \infty$
\Statex{}
\Statex {\bf \underline {begin}}
\State Form the matrix $\bar{\bf V}=
\left[
\begin{array}{c|c|c}
{\bf V}^T & \cdots & {\bf V}^T\\
\end{array}
\right]^T$ by repeating $\bf V$, $2(\ceil{\psi}+1)$ times. 
\State Form the vector $\bar{\bf c}=
\left[
\begin{array}{c|c|c}
{\bf c}_{L_1}^T & \cdots & {\bf c}_{L_2}^T\\
\end{array}
\right]^T$ where $L_1 = -\ceil{\psi}$ and $L_2 = \ceil{\psi}+1$ and ${\bf c}_i = (i-\frac{1}{2} + \floor{{\bf A}^{-1}{\bf y}})\mathbf{1} - {\bf A}^T{\bf y}$ and $\mathbf{1}$ is of length $n$.
\For {$i=1$ to $2n(\ceil{\psi}+1)$}
\If {$\bar{c}_i > 0$}
\State $\bar{c}_i \leftarrow -\bar{c}_i $.
\State $\bar{V}_{\{i\}} \leftarrow -\bar{V}_{\{i\}} $.
\EndIf
\EndFor
\State $\Phi =$ CellEnum($\bar{\bf V},{\bar{\bf c}},\{+,\cdots,+\}$)
\For {each ${\bf d}\in \Phi$}
\State Find ${\bf a} = \round{{\bf V}{\bf d} + {\bf A}^T{\bf y}}$
\If {$f({\bf a})< f_{\text{min}}$}
\State Set ${\bf a}^* = {\bf a}$.
\State Set $f_{\text{min}} = f({\bf a})$. 
\EndIf
\EndFor

\end{algorithmic}
\caption[caption]{CVP for IP$^k$ matrices}
\label{cvpExact}
\end{algorithm}

These modifications do not affect the complexity order of the algorithm. Equation \eqref{eqn:complexity} still describes the complexity except for the fact that now $\psi = \sqrt{\frac{G_{max}}{\lambda_{min}}}$. 

The results can be extended to DP$^k$ matrices. In this case, the optimal coefficient vector is given by

\begin{theorem}
The solution to the CVP for DP$^k$ matrices satisfies:
\begin{equation}
{\bf a}^* = \left\lceil{{\bf D}^{-1}({\bf V}{{\bf x + A}^T{\bf y}})}\right\rfloor
\label{eqn:round_dpk}
\end{equation}
for some ${\bf x}\in \mathbb{R}^k$. Furthermore, 
\begin{equation}
\|{\bf a}^* - {\bf A}^{-1}{\bf y}\| \le \psi = \sqrt{\frac{{G}_{max}}{\lambda_{min}}}.
\label{eqn:bound}
\end{equation}
\label{thm:cvpthm_dpk}
\end{theorem}

We will now propose a generalization of our approximation algorithm for the CVP.  First note that for the positive-definite matrix ${\bf G}$ with normalized and sorted eigenvalues $\lambda_i$ we have ${\bf G} = {\bf A}^T{\bf A}$
 if and only if 
 \begin{equation}
 {\bf A =UB}
 \label{eqn:ubd}
 \end{equation}
 where ${\bf U}$ is an arbitrary unitary matrix and ${\bf B}$ is a symmetric matrix with the same eigenvectors as ${\bf G}$ and with eigenvalues $\beta_{i}$ (sorted) that satisfy:
$$\beta_i^2 = \lambda_i.$$
Without loss of generality, we can also assume $\beta_i$'s are positive (if they are negative, we can transfer the sign to the unitary matrix $\bf U$). Under this assumption, if ${\bf G}$ is $\text{IP}^k$ then the matrix ${\bf B}$ must be $\text{IP}^k$ too.
Finally, we can generalize Theorem \ref{thm:mainthm} for CVP. Here, besides mapping the matrix $\tilde{\bf G}$ to its IP$^k$ approximation, we will also need to map the vector ${\bf y}$, whose nearest neighbor is of interest, to a different vector in space. 

\begin{theorem}
Let ${\bf y}$ be an arbitrary vector in $\mathbb{R}^n$ and $\tilde{\bf A}\in \mathbb{R}^{n\times n}$ be a full-rank matrix satisfying $\tilde{\bf A} = {\bf U}\tilde{\bf B}$ as in Equation \eqref{eqn:ubd}. Furthermore assume $\tilde{\bf B}^T\tilde{\bf B}$ is $\widetilde{\text{IP}}^k_\gamma$. Define 
$$f({\bf a}) = \|\tilde{\bf A}{\bf a-y}\|^2.$$
Suppose ${\bf a}^*$ is the solution to 
\begin{equation*}
{\bf a}^* = \argmin_{{\bf a\in Z}^n}f({\bf a})
\end{equation*}
and $\hat{\bf a}$ satisfies
\begin{equation*}
\hat{{\bf a}} = \argmin_{{\bf  a\in Z}^n}\|\hat{\bf A}{\bf a}-\hat{\bf y}\|^2
\end{equation*}
where $\hat{\bf A} = \mathcal{I}_k(\tilde{\bf B})$ and $\hat{\bf y} = \hat{\bf A}\tilde{\bf A}^{-1}{\bf y}$.
we have that:
\begin{equation*}
f(\hat{\bf a}) < \frac{1}{1-\gamma} f({\bf a}^*).
\end{equation*}
\label{thm:mainthmcvp}
\end{theorem}

To summarize, we propose Algorithm \ref{approxCVP} for approximating the CVP for $\widetilde{\text{IP}}^k_\gamma$ matrices.

\begin{algorithm}[H]
\caption[caption]{$\frac{1}{1-\gamma}$-approximation algorithm for the CVP for $\widetilde{\text{IP}}^k_\gamma$ matrices}
\begin{algorithmic}[1]
\Statex {{\bf Input:} The vector ${\bf y}$, the full-rank square matrix ${\tilde{\bf A} ={\bf U}\tilde{\bf B}}$ as in Equation \eqref{eqn:ubd} with $\tilde{\bf B}^T\tilde{\bf B}$ being $\widetilde{\text{IP}}^k_\gamma$}.
 \State Find $\hat{\bf A} = \mathcal{I}_k(\tilde{\bf B})$ and $\hat{\bf y} = \hat{\bf A}\tilde{\bf A}^{-1}{\bf y}$.
\State \Return {$\hat{\bf a}$ , the output of Algorithm \ref{cvpExact} applied on $\hat{\bf y}$ and $\hat{\bf A}$.}
\end{algorithmic}
\label{approxCVP}
\end{algorithm}
Again, the complexity follows Equation \eqref{eqn:complexity} where $\psi$ is equal to $\sqrt{\frac{G_{max}}{\lambda_{min}}}$, the ratio of the largest diagonal element and the smallest eigenvalue of the matrix $\hat{\bf G}$

\begin{rem}
A similar approximation factor of $\frac{1}{1-\gamma}$ can be obtained for a general positive-definite matrix of the form ${\bf G} = \sqrt{\bf D}({\bf I} - {\bf P})\sqrt{\bf D}$ if we instead solve the SVP or CVP for the matrix  $\hat{\bf G} = \sqrt{\bf D}\mathcal{I}_k({\bf I} - {\bf P})\sqrt{\bf D}$. 
\end{rem}

\subsection{Application: MIMO Detection; a Trade-off Between Complexity and Accuracy}
We study a potential application of Algorithm \ref{approxCVP} in the context of MIMO detection. Consider communication over a general MIMO channel without CSIT: 
\begin{equation*}
{\bf y = Hx + z} .
\end{equation*}
Assume the noise vector ${\bf z}$ is i.i.d. Gaussian. The receiver performs ML detection to estimate ${\bf x}$. After shifting and scaling, and assuming the alphabet size is large enough, we get the following optimization problem \cite{damen2003maximum}: 
\begin{equation*}
\hat{\bf x} = \argmin_{{\bf x}\in\mathbb{Z}^n} \|{\bar{\bf y}} - \bar{\bf H}{\bf x}\|^2.
\end{equation*}
This can be seen as an instance of the CVP, where the lattice matrix is $\bar{\bf H}$. Note that the ML detector outputs the correct value of ${\bf x}$, if the norm of the equivalent noise vector $\bar{\bf z}$ is smaller than half the minimum distance of the vectors (or equivalently, half the length of the shortest vector) of the lattice characterized by $\bar{\bf H}$. Let us call this parameter $d^H_{min}$. Thus we have:

\begin{equation}
P_{er} \le P (\|\bar{\bf z}\|> d^H_{min}/2)
\label{eqn:MLsolution}
\end{equation}

which can be expressed in terms of the CCDF of the Chi-squared distribution. 
Let us define $\lambda^H_{max}$ as the maximum eigenvalue of $\bar{\bf H}$. For any integer $k = 0,\dots,n$, suppose $\gamma(k)$ is the smallest positive number for which the lattice $\frac{1}{\lambda^H_{max}}\bar{\bf H}$ is $\widetilde{\text{IP}}^k_{\gamma(k)}$. 
We will show that Algorithm \ref{approxCVP} achieves the following error probability:

\begin{equation}
P^{alg5}_{er} \le P \left(\|\bar{\bf z}\|> \frac{d^H_{min}}{1+1/\sqrt{1-\gamma(k)}}\right). 
\label{eqn:IPKsolution}
\end{equation}

Note that if for some integer $k$ we have $\gamma(k) = 0$, the matrix $\frac{1}{\lambda^H_{max}}\bar{\bf H}$ will be $\text{IP}^k$, and not surprisingly the algorithm returns the ML solution which achieves the same error probability as in Equation \eqref{eqn:MLsolution}. But in general there will be a trade off between the complexity of the decoder and the achievable error probability: as we let $k$ decrease to zero, Algorithm \ref{approxCVP} runs faster (as evident from Equation \eqref{eqn:complexity}) but $\gamma(k)$ becomes larger which indicates a higher error probability, according to Equation \eqref{eqn:IPKsolution}.\\
To prove Equation \eqref{eqn:IPKsolution}, let us suppose that $\|\bar{\bf z}\|< \frac{d^H_{min}}{1+1/\sqrt{1-\gamma(k)}}$. Assume $\hat{\bf x}$ is the output of Algorithm \ref{approxCVP} applied on  $\frac{1}{\lambda^H_{max}}\bar{\bf y}$ and $\frac{1}{\lambda^H_{max}}\bar{\bf H}$. If $\hat{\bf x}\neq {\bf x}$ , the best achievable approximation factor is:
\begin{eqnarray*}
\left(\frac{\|\bar{\bf H}\hat{\bf x} - \bar{\bf y}\|}{\|\bar{\bf H}{\bf x} - \bar{\bf y}\|}\right)^2 & \ge& \left(\frac{d^H_{min}-\|\bar{\bf z}\|}{\|\bar{\bf z}\|}\right)^2
\\&>& \frac{d^H_{min} - \frac{d^H_{min}}{1+1/\sqrt{1-\gamma(k)}}}{ \frac{d^H_{min}}{1+1/\sqrt{1-\gamma(k)}}}\\
&=& \left(\frac{1}{\sqrt{1-\gamma(k)}}\right)^2 = \frac{1}{1-\gamma(k)}.
\end{eqnarray*}    
This contradicts with the fact that Algorithm \ref{approxCVP} achieves an $\frac{1}{1-\gamma(k)}$ approximation factor. Thus we must have that  $\hat{\bf x}= {\bf x}$. As a result, as long as $\|\bar{\bf z}\|< \frac{d^H_{min}}{1+1/\sqrt{1-\gamma(k)}}$ , Algorithm 5 outputs the correct value of ${\bf x}$. This proves Equation \eqref{eqn:IPKsolution}.

\section{Open Problem: $\widetilde{\text{IP}}^k$-reduced Basis}
\label{sec:reduction}
The basis matrix of a lattice is not unique. For any lattice $\mathcal{L}({\bf A})$ there are infinitely many bases. All these bases are related via linear transformation by unimodular matrices. In other words if we have $\mathcal{L}({\bf A}) = \mathcal{L}({\bf B})$ then there exists a unimodular matrix ${\bf T}$ such that ${\bf B = AT}$.

The field of lattice reduction aims at finding such unimodular transformations for arbitrary lattice bases, and producing new bases with more desirable properties. The new basis is usually called a reduced basis of the lattice. There are different notions of lattice reduction. For instance Minkowski's criteria for calling a basis ${\bf A}$ reduced is that the shortest vector (or column)  of this basis, ${\bf v}_1$ must be the shortest vector of the lattice $\mathcal{L}({\bf A})$; the second shortest vector of ${\bf A}$ must be the second shortest vector of the lattice among all the vectors that are linearly independent of ${\bf v}_1$ and so on. Of course, there is no polynomial time algorithm known that can find such a reduced basis (as otherwise, the SVP would have been solved and much more). Another notion is the LLL-reduced basis due to \cite{lenstra1982factoring}. An LLL reduced basis can be found in polynomial time. However the shortest vector of an LLL-reduced basis can be exponentially longer than the shortest vector of the lattice.

\begin{figure}[H]
\begin{center}
\includegraphics[height = 7 cm]{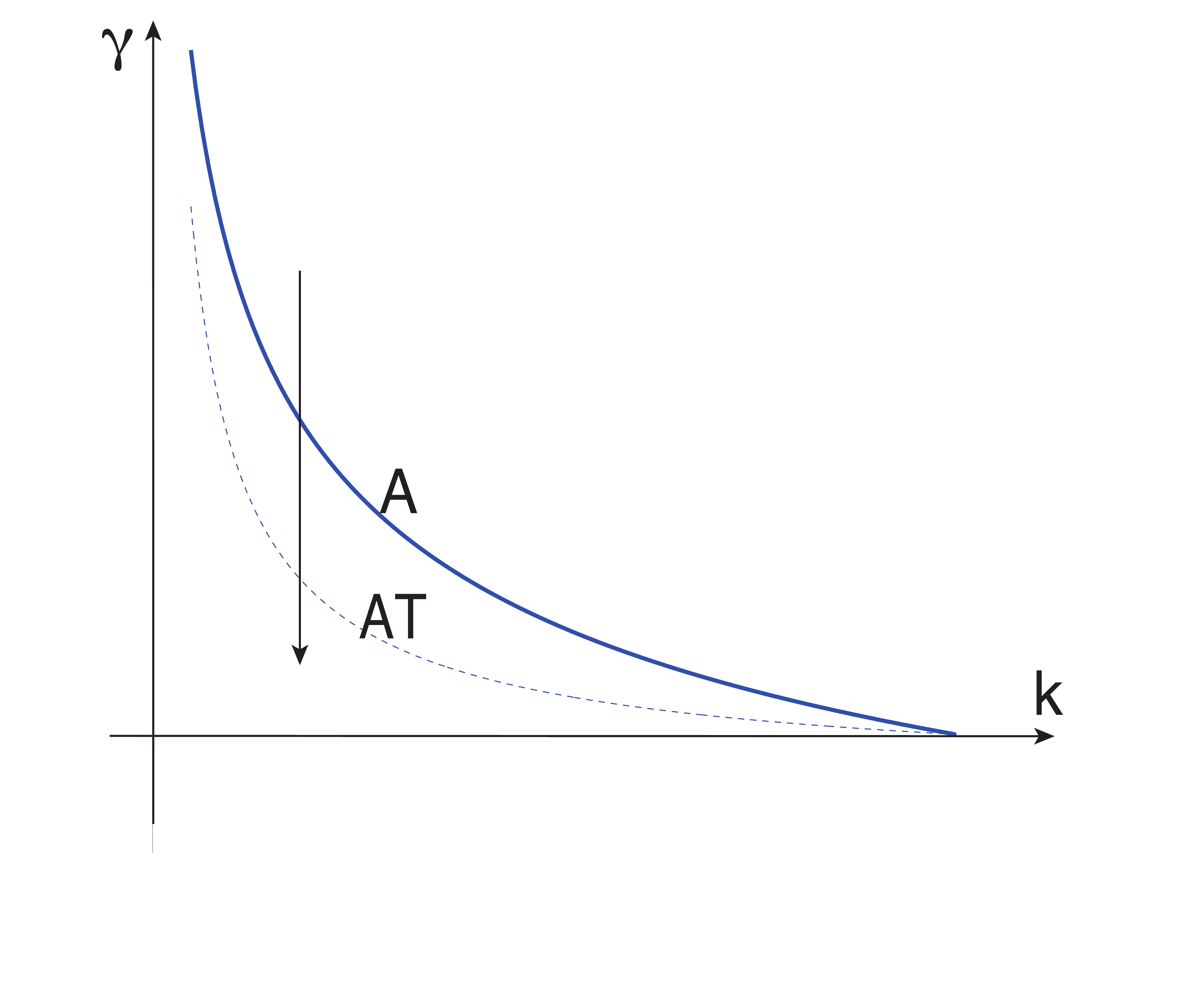}
\end{center}
\vspace*{-1 cm}\caption[caption]{An arbitrary lattice basis and its $\widetilde{\text{IP}}^k$ reduction}
\label{fig:reduced}
\end{figure}

Here, based on the concept of $\widetilde{\text{IP}}_\gamma^k$ matrices we introduce a new notion of reduced basis. Specifically, we define
\begin{defn}
A lattice basis ${\bf A}$ is called $\widetilde{\text{IP}}^k$-reduced if it holds that $\gamma_{\bf A}(k) \le \gamma_{\bf B}(k)$ for any matrix $\bf B$ that satisfies $\mathcal{L}({\bf B}) = \mathcal{L}({\bf A})$. Here $\gamma_{\bf A}(k)$ is the smallest $\gamma$ for which the matrix ${\bf G} ={\bf A}^T{\bf A}$ (after normalizing the eigenvalues) is $\widetilde{\text{IP}}^k_\gamma$.
\end{defn}
In other words, given an arbitrary lattice basis, we are interested in finding a new basis for the same lattice which minimizes the value of ${\bf \gamma}$ for a particular $k$. This is demonstrated in Figure \ref{fig:reduced}. 

It should be clear why we are interested in such a basis. We want to achieve the best possible approximation factor for SVP and CVP through Algorithms \ref{approxSVP} and \ref{approxCVP} which run in complexity order of $O(n^{k+1}(2\ceil{\psi}+2)^{k+1})$. Currently we do not know any algorithm that can find an $\widetilde{\text{IP}}_\gamma^k$-reduced basis for an arbitrary lattice. Finding an efficient algorithm which performs this task could have quite interesting implications in terms of approximating the SVP or CVP for a general lattice. 
\section{Appendices}
\label{sec:Appendices}
\subsection{Proof of Theorem \ref{thm:cvpthm_dpk}}
\label{subsec:cvpthm}
\begin{IEEEproof}
We will prove the claim for DP$^k$ matrices, that is ${\bf G } = {\bf D} - {\bf P} = {\bf D} - {\bf VV}^T$. Theorem \ref{thm:cvpthm} follows as a special case. To simplify the notation, we define ${\bf z = A}^T{\bf y}$. We can rewrite $f({\bf a}) = {\bf a^TGa} -2 {\bf z}^T{\bf a} + {\bf y}^T{\bf y}$ as follows:
$$f({\bf a}) = \sum_{i=1}^n (D_{ii}-P_{ii})a_i^{2} - 2\sum_{i=1}^n\sum_{j=1}^{i-1}P_{ij}a_ia_j -2\sum_{i=1}^n z_ia_i +  {\bf y}^T{\bf y}.$$
First note that since ${\bf P}$ is positive semi-definite, we have $P_{ii}\ge 0$ for $i=1,\dots,n$. If for some $j$ we have $P_{jj} = 0$, then we must have $P_{ij} = P_{ji} = 0$ and $V_{ji}= 0$ for $i=1,\dots,n$. The optimal value for $a_j$ in this case is simply $a_j^*=\left\lceil\frac{z_j}{D_{jj} }\right\rfloor$ which satisfies the claim of the theorem. The problem can then be reduced to $n-1$ dimensions. Thus without loss of generality we assume $P_{ii}>0$ for the rest of the proof.\\
Assume that we already know the optimal value for all $a_i^*$ elements except for one element, $a_j$. Note that $f$ is a convex parabola in $a_j$ (this is because $D_{jj}-P_{jj} = G_{jj}$ is a diagonal element of a positive-definite matrix) thus the optimal integer value for $a_j$ is the closest integer to its optimal real value. By taking partial derivative with respect to $a_j$, the optimal real value of $a_j$ is easily seen to be equal to 
$$\frac{z_j+\sum_{i=1, i\neq j}^nP_{ij}a_i^*}{D_{jj} - P_{jj}}.$$
Taking the closest integer to the real valued solution, we find:
\begin{equation}
\begin{aligned}
\Rightarrow a_j^*&=\left\lceil\frac{z_j+\sum_{i=1, i\neq j}^nP_{ij}a_i^*}{D_{jj} - P_{jj}}\right\rfloor \;,\; or\\
a_j^*&=\left\lfloor\frac{z_j+\sum_{i=1, i\neq j}^nP_{ij}a_i^*}{D_{jj} - P_{jj}}\right\rceil.
\label{eqn:firstineq}
\end{aligned}
\end{equation}
Due to the symmetry of the parabola, both functions return equally correct solutions for $a_j^*$. \\
Note that this expression must be true for any $j$: If for ${\bf a^*}$ and for some $j$, $a_j^*$ does not satisfy at least one of these two equations, we can achieve a strictly smaller value over $f$ by replacing $a_j^*$ with the value given above, and so ${\bf a}^*$ cannot be optimal. From Equation \eqref{eqn:firstineq} we have that:
\begin{align}
a_j^*+\frac{1}{2}&\ge \frac{z_j+\sum_{i=1, i\neq j}^nP_{ij}a_i^*}{D_{jj} - P_{jj}} \;\;\;,\;\;and \label{eqn:line1}\\
a_j^*-\frac{1}{2}&\le \frac{z_j+\sum_{i=1, i\neq j}^nP_{ij}a_i^*}{D_{jj} - P_{jj}}\label{eqn:line2}.
\end{align}
Starting with Equation \eqref{eqn:line1}, we multiply both sides by the denominator, and add the term $a_j^*P_{jj}$ to obtain:
\begin{align*}
(a_j^*+\frac{1}{2})D_{jj}&\ge z_j+\sum_{i=1}^nP_{ij}a_i^*+\frac{1}{2}P_{jj}.\\
\end{align*}
Dropping the positive term $\frac{1}{2}P_{jj}$ we conclude 
$$(a_j^*+\frac{1}{2})D_{jj}> z_j+\sum_{i=1}^nP_{ij}a_i^*.$$
As a result, we have:
\begin{align*}
(a_j^*+\frac{1}{2})D_{jj}&> z_j+\sum_{i=1}^nP_{ij}a_i^*\\
\Rightarrow (a_j^*+\frac{1}{2})&> \frac{z_j+\sum_{i=1}^nP_{ij}a_i^*}{D_{jj}} \; , \; j=1\dots n.
\end{align*}
Writing this inequality in vector format, we obtain
\begin{equation}
{\bf a}^*+\frac{1}{2}\mathbf{1}> {\bf D}^{-1}({\bf z}+{\bf P}^T{\bf a}^*) = {\bf D}^{-1}({\bf z}+{\bf V}({\bf V}^T{\bf a}^*))
\label{eqn:IV}
\end{equation}
In a similar fashion one can show that Equation \eqref{eqn:line2} results in
\begin{equation}
 {\bf a}^*-\frac{1}{2}\mathbf{1}< {\bf D}^{-1}({\bf z}+{\bf V}({\bf V}^T{\bf a}^*)).
\label{eqn:V}
\end{equation}
Defining ${\bf x} = {\bf V}^T{\bf a}^*$, it follows from \eqref{eqn:IV} and \eqref{eqn:V} that
$${\bf a^*} - \frac{1}{2}\mathbf{1}< {\bf D^{-1}(Vx+z)} < {\bf a^*} + \frac{1}{2}\mathbf{1}$$
$$\Rightarrow {\bf a}^* = \round{{\bf D^{-1}(Vx+z)}} = \round{{\bf D}^{-1}({\bf Vx}+{\bf A}^T{\bf y})}.$$
This completes the proof of Equation \eqref{eqn:round}.\\
To prove Equation \eqref{eqn:bound}, note that  
$$\|{\bf A}{\bf a}^* -{\bf y}\|^2 = \|{\bf A}({\bf a}^* - {\bf A}^{-1}{\bf y})\|^2 \ge \lambda_{min}\|{\bf a}^* - {\bf A}^{-1}{\bf y}\|^2.$$
It is also evident that $\|{\bf A}{\bf a}^* -{\bf y}\|^2\le G_{max}$, that is the square distance of the closest vector of the lattice to ${\bf y}$ is less than the largest diagonal element of ${\bf G}$. To see this, note that for any point in the Voronoi region of the origin there are $N$ successive minima of the lattice ($\nu_1 , \dots , \nu_N$) such that $p = \sum_{i=1}^N{\theta_i\nu_i}$ where $\theta_i\ge0$ and $\sum_{i=1}^N\theta_i \le 1$. The claim follows since $\sqrt{G_{max}}$ cannot be smaller than the length of the $N$'th successive minima of the lattice. 
$$G_{max}\ge \|{\bf A}{\bf a}^* -{\bf y}\|^2 \ge \lambda_{min}\|{\bf a}^* - {\bf A}^{-1}{\bf y}\|^2 $$
from which we can conclude: $\|{\bf a}^* - {\bf A}^{-1}{\bf y}\| \le\sqrt{\frac{{G}_{max}}{\lambda_{min}}}.$ 
\end{IEEEproof}

\subsection{Proof of Theorem \ref{thm:mainthm}}
\label{sec:prooflemma2}
\begin{IEEEproof}
Proof of Theorem \ref{thm:mainthm} can be seen as a special case of proof of Theorem \ref{thm:mainthmcvp} for ${\bf y = 0}.$ The extra condition of ${\bf a\neq 0}$ for the SVP does not cause any problem here.
\end{IEEEproof}

\subsection{Proof of Theorem \ref{thm_dpk}}
\label{subsec:mainthmsvp}
\begin{IEEEproof}
The proof of Theorem \ref{thm_dpk} is almost identical to that of Theorem \ref{thm:cvpthm_dpk} after setting ${\bf y} = {\bf 0}$. The requirement for checking the standard unit vectors follows from the fact that ${\bf a}$ cannot be the all zero vector. In other words, all the elements of the vector ${\bf a}$ must satisfy \eqref{eqn:firstineq} unless ${\bf a}^*$ is a standard unit vector in the direction of the $j$'th axis in which case replacing $a_j^*$ by  \eqref{eqn:firstineq} will lead to the all zero vector.

To prove the bound on norm of ${\bf a}^*$ note that the square norm of the shortest vector of the lattice must be smaller or equal to $G_{min}$. Therefore, 
\begin{equation*}
G_{min}\ge \|{\bf A}{\bf a}^*\|^2 \ge \lambda_{min}\|{\bf a}^*\|^2 \rightarrow \|{\bf a}^*\| \le\sqrt{\frac{{G}_{min}}{\lambda_{min}}}.
\end{equation*}

\end{IEEEproof}

\subsection{Proof of Theorem \ref{thm1_dp1}}
\label{subsec:ip1proof}
\begin{IEEEproof}
This is clearly a special case of Theorem \ref{thm_dpk}. The fact that we can limit the search to $x\in \mathbb{R}^+$ instead of the whole $\mathbb{R}$ follows trivially from: ${\bf a}^T{\bf G}{\bf a} = ({-\bf a})^T{\bf G}({-\bf a})$.
\end{IEEEproof}

\subsection{Proof of Theorem \ref{thm:mainthmcvp}}
\label{sec:prooflemma}
\begin{IEEEproof}
We prove the claim for the more general setting where $\tilde{\bf A} = {\bf U}\tilde{\bf B}\sqrt{\bf D}$ where ${\bf D}$ is diagonal with strictly positive diagonal elements. First define ${\bf b}^* = {\bf Da}^{*}$ and $\hat{\bf b} = {\bf D}\hat{\bf a}$ and $\hat{\bf B} = \mathcal{I}_k(\tilde{\bf B})$ and ${\bf c} = \tilde{\bf B}^{-1}{\bf U}^T{\bf y}$. We have:
\begin{equation*}
f({\bf a}^*)=\|\tilde{\bf B}\sqrt{\bf D}{\bf a}^* -\tilde{\bf B}\sqrt{\bf D}\sqrt{\bf D}^{-1}\tilde{\bf B}^{-1}{\bf U}^T{\bf  y}\|^2
\end{equation*}
which can be rewritten as 
\begin{eqnarray*}
&f({\bf a}^*)= \|\tilde{\bf B}({\bf b}^* - {\bf c})\|^2\\
&= \|{\bf b}^{*} -{\bf c}\|^2 - \sum_{i=1}^n(1-\beta_i^2)(({\bf b}^* -{\bf c})^T {\bf v}_i)^2
\end{eqnarray*}
where $\beta_i$'s and $ {\bf v}_i$'s are the eigenvalues (sorted) and eigenvectors of $\tilde{\bf B}$. That is,
\begin{equation*}
\tilde{\bf B}= {\bf I} - \sum_{i=1}^n(1-\beta_i){\bf v}_i{\bf v}_i^T.
\end{equation*}
For the vector ${\bf b}^* - {\bf c}$ we have:
\begin{equation}
\|{\bf b}^*- {\bf c}\|^2 \ge \sum_{i=1}^n (({\bf b}^* - {\bf c})^T {\bf v}_i)^2.
\label{eqn:proof1}
\end{equation}
This directly follows from the fact that ${\bf v}_i$'s are orthonormal. From Equation \eqref{eqn:proof1} we conclude:
\begin{eqnarray*}
&(1-\gamma)\|{\bf b}^*- {\bf c}\|^2\ge \|{\bf b}^*- {\bf c}\|^2 -\gamma \sum_{i=1}^{n}(({\bf b}^* - {\bf c})^T {\bf v}_i)^2.\end{eqnarray*}Therefore,
\begin{eqnarray*}&(1-\gamma)\|{\bf b}^*- {\bf c}\|^2 - \sum_{i=1}^{k}(1-\beta_i^2-\gamma)(({\bf b}^*- {\bf c})^T {\bf v}_i)^2 \ge\\
&\|{\bf b}^*- {\bf c}\|^2 -\sum_{i=1}^k(1-\beta_i^2) (({\bf b}^* - {\bf c})^T {\bf v}_i)^2-\\
&\sum_{i=k+1}^n\gamma (({\bf b}^* - {\bf c})^T {\bf v}_i)^2\overset{\#}{\ge}\\
&\|{\bf b}^*- {\bf c}\|^2 -\sum_{i=1}^n(1-\beta_i^2) (({\bf b}^* - {\bf c})^T {\bf v}_i)^2
\end{eqnarray*}
where ($\#$) follows from the fact that $\gamma \le 1- \beta_i^2$ for $i=k+1,\dots,n$. 
Thus we can bound $f({\bf a}^*)$ as follows
\small
\begin{eqnarray*}
f({\bf a}^*)  \hspace{-0.2cm}&=&\hspace{-0.2cm}  \|\tilde{\bf B}({\bf b}^* - {\bf c})\|^2\\
&=&\hspace{-0.2cm}  \|{\bf b}^*- {\bf c}\|^2 - \sum_{i=1}^n(1-\beta_i^2)(({\bf b}^* - {\bf c})^T\cdot {\bf v}_i)^2\\
&\overset{+}{\ge}&\hspace{-0.2cm}(1-\gamma)\|{\bf b}^*- {\bf c}\|^2 - \sum_{i=1}^{k}(1-\beta_i^2-\gamma)(({\bf b}^*- {\bf c})^T {\bf v}_i)^2 \\
&=&\hspace{-0.2cm} (1-\gamma)(\|{\bf b}^* - {\bf c}\|^2 - \sum_{i=1}^{k}\frac{1-\beta_i^2-\gamma}{1-\gamma}(({\bf b}^* - {\bf c})^T\cdot {\bf v}_i)^2)\\
&\overset{\#}{\ge}&\hspace{-0.2cm}(1-\gamma)(\|{\bf b}^* - {\bf c}\|^2 - \sum_{i=1}^{k}(1-\beta_i^2)(({\bf b}^* - {\bf c})^T\cdot {\bf v}_i)^2)\\
&=&\hspace{-0.2cm}(1-\gamma)\|{\hat{\bf B}({\bf b}^{*} - {\bf c})}\|^2\\
&=&\hspace{-0.2cm}(1-\gamma)\|\hat{\bf A}{\bf a}^{*} - \hat{\bf y}\|^2\\
&\ge&\hspace{-0.2cm}(1-\gamma)\|\hat{\bf A}\hat{\bf a} - \hat{\bf y}\|^2\\
&\overset{**}{\ge}&\hspace{-0.2cm}(1-\gamma)f(\hat{\bf a})
\end{eqnarray*}\normalsize
where ($\#$) follows from $\beta_i^2 \le 1$ and $\gamma < 1$ and ($**$) is due to the fact that $\|\tilde{\bf A}{\bf a} - {\bf y}\| \le \|\hat{\bf A}{\bf a} - \hat{\bf y}\|$ for any arbitrary vector ${\bf a}$. 
\end{IEEEproof}

\bibliographystyle{IEEEtran}
\bibliography{IEEEfull,main}

\end{document}